# AgBiSe$_2$ Colloidal Nanocrystals for Use in Solar Cells

*M. Zafer Akgul[a] and Gerasimos Konstantatos[a,b]\**

[a]ICFO-Institut de Ciències Fotòniques, The Barcelona Institute of Science and Technology, 08860 Castelldefels (Barcelona), Spain

[b]ICREA—Institució Catalana de Recerca i Estudis Avançats, Passeig Lluís Companys 23, 08010 Barcelona, Spain

E-mail: gerasimos.konstantatos@icfo.eu

**Abstract**

Metal selenide nanocrystals have attracted attention as promising materials in photovoltaics and thermoelectrics. However, the expensive and labor-intensive synthesis methods utilized for the production of these nanomaterials have impeded their widespread utilization. The need for air-free environment and high synthesis temperature for crystal nucleation and growth lead as the major factors contributing to the cost of synthesis. In this work, we present a synthesis method for metal selenide nanocrystals at room temperature under ambient conditions that is enabled by a cost-effective selenium precursor. Thanks to the relative abundance and low toxicity, ternary silver bismuth selenide (AgBiSe$_2$) is used as the prototyping material, as well as silver bismuth sulfide (AgBiS$_2$) and alloyed silver bismuth sulfide-selenide (AgBiS$_x$Se$_y$) to show the bandgap tunability of the obtained nanocrystals by utilizing a simple mixed anion precursor approach. A preliminary solar cell made of colloidal AgBiSe$_2$ nanocrystals synthesized via the proposed ambient condition method yields a power conversion efficiency up to 2.6 %, which is the first colloidal AgBiSe$_2$ nanocrystal solar cell reported in the literature to the best of author's knowledge. This synthesis route is expected to pave the way for low-cost environmentally friendly solution-processed photovoltaics.

KEYWORDS: (non-toxic, room temperature, semiconductors, nanocrystals, solar cells)

**Introduction**

Due to the endlessly increasing energy consumption, cost-effective alternative energy sources have been put under investigation. To this purpose, clean and renewable energy sources, such as solution-processed photovoltaics and thermoelectrics, have recently been shown to hold great promise thanks to their advantage in low manufacturing cost and scalability. However, the narrow bandgap materials used in these applications are typically based on heavy metal chalcogenides (such as PbS).[1,2] Due to the inherent toxicity and difficulty in synthesis of these materials, the expected jump in the development of low-cost photovoltaics and thermoelectrics has not taken place, yet. Although some attempts have been made to eliminate the costly synthesis conditions for the metal selenide semiconductors partially or completely, a limited success has been achieved due to the necessity to employ inert reaction environment and high temperature selenization.[3,4] To our knowledge, no direct synthesis method for non-toxic and earth-abundant metal selenide colloidal nanocrystals (CNCs) entirely under ambient conditions and at room temperature has been demonstrated so far.

One of the major issues with the synthesis of metal selenide CNCs is the availability of an air-stable, cost-effective and sufficiently reactive selenium precursor to replace the combination of elemental selenium-high reaction temperature utilized in the existing reports.[5–7] Despite alternative selenium precursors have been proposed in the literature, such as sodium hydroselenide (NaHSe), bis (trimethylsilyl) selenide ([Me$_3$Si]$_2$Se), trioctylphosphine-Se (TOP-Se) and tributylphoshine-Se (TBP-Se),[8–13] no commercially-viable solution for large-scale production of metal selenide CNCs has been demonstrated so far because of the air sensitivity and/or high cost associated with alkylphosphines. Apart from complexes of elemental Se with TOP and TBP, other strategies have been put into use to prepare a viable selenium precursor using elemental selenium. As the most prominent examples, the solution of elemental Se in octadecene (ODE), oleic acid (OA) and/or oleylamine (OLA) has been utilized to eliminate the problems related to alkyl phosphines thanks to their lower cost and better stability under ambient conditions.[14–17]

In addition, alky ammonium (poly)selenides have been found to be promising thanks to their relatively low cost and high solubility in alkyl amines. One of the early reports on this route used a strong reducing agent, sodium borohydrate ($NaBH_4$), for the reduction of elemental selenium into selenide ions to form the alkyl ammonium (poly)selenides.[18] While colloidal nanocrystals were produced successfully using this Se precursor, the necessity of air-free synthesis techniques remained in place because of the air sensitivity of $NaBH_4$. Furthermore, the possibility of Na-doping of the nanocrystals remained as an open question to be answered due to the byproducts of reaction remaining in the synthesis mixture after the formation of alkyl ammonium (poly)selenides. Recently, it has been demonstrated that the dissolution of elemental selenium in a mixture of oleylamine and alkyl thiols can occur readily without heating.[4] In the given reaction, it was reported that the alkyl thiols could effectively reduce elemental selenium to selenide anions in the presence of oleylamine and that the oleylamine molecules could bind to selenide ions to form alkyl ammonium (poly)selenide compounds. This strategy has attracted significant attention thanks to the ease and effectiveness of amine-assisted Se dissolution in organic solvents and the absence of foreign cations originating from the use of inorganic reducing agents, such as $NaBH_4$. One of the key findings of this study was that alkyl ammonium (poly)selenides could stay intact under ambient conditions for sufficiently long times if excess thiol was provided into the solution.[4]

It is advantageous to employ a reaction path that can utilize an air-stable Se source as the starting material to eliminate the need for protection of the Se source from degradation. In this work, selenium dioxide ($SeO_2$) is chosen as the selenium source instead of elemental selenium as elemental selenium is typically prone to oxidation in air forming $SeO_2$. By employing $SeO_2$, we extend the results on the alkyl ammonium (poly)selenides precursor and show that the mixed amine-thiol solvent is capable of reducing not only elemental selenium ($Se^0$) as shown by a previous report,[4] but also $SeO_2$ ($Se^{4+}$) into selenide ions, which enables a facile, fast and low-cost synthetic method using only air-stable and commercially available chemicals. We demonstrate that the obtained selenium precursor is sufficiently reactive towards silver and bismuth iodide salts at room temperature to form silver bismuth selenide ($AgBiSe_2$) CNCs without requiring a chemically inert reaction environment or heating, unlike the results reported previously. This, in turn, enables the utilization of a much-simplified synthesis setup compared to the hot-injection methods that utilize Schlenk line setup. Thanks to the strong antioxidant property of thiol-amine mixture on this selenide precursor, it is demonstrated that various binary, ternary and quaternary metal selenide CNCs can be obtained within minutes.

**Experimental Section**

The details of the synthesis and of the characterization methods used in this study are given in the supporting information (SI).

**Results and Discussion**

In this study, we make use of $AgBiSe_2$, a member of I-III-$VI_2$ ternary metal chalcogenide semiconductor family, as a model system to demonstrate the versatility of the Se precursor prepared with $SeO_2$ and amine-thiol solvent mixture. $AgBiSe_2$ is chosen because it combines low toxicity and high abundance of the constituent elements with attractive optical and electronic properties (for example a bandgap of 0.6 eV in bulk[19] and 0.7 - 1.0 eV in nanocrystalline form[20,21]) on par with other materials that are considered important for photovoltaics and thermoelectrics, such as $CuInSe_2$[17] and $AgInSe_2$[22] and PbSe.[23] Although the binary compounds of the constituent elements, namely silver selenide ($Ag_2Se$) and bismuth selenide ($Bi_2Se_3$), have been studied extensively in the literature thanks to their attractive optical and thermoelectric properties, the number of reports dealing with this ternary semiconductor is limited.[24–30]

$AgBiSe_2$ has recently attracted attention for its potential applications, especially in thermoelectrics.[19,31,32] Despite the attention for its promising performance in thermoelectrics, the studies mainly focused on either bulk or in thin films of $AgBiSe_2$.[33–36] Hitherto only a few reports have been dedicated to the synthesis of colloidal $AgBiSe_2$ nanocrystals and these were performed under a controlled environment to protect the precursors from oxidation.[20,37] This results from the use of either elemental selenium that necessitates high reaction temperature or more reactive precursors, such as $Bi[(N(SiMe_3)_2]_3$ that are also prone to oxidation under ambient conditions.[20,37] Thus, addressing the oxidation problem of precursors while keeping the reaction temperature as close to room temperature as possible for a low-cost ambient condition synthesis of $AgBiSe_2$ CNCs still remains as a challenge.

In the previous studies, oleylamine was used extensively as the primary alky amine to dissolve elemental selenium via formation of oleylammonium (poly)selenides. However, it is known that the reactivity of a precursor generally decreases if longer groups are attached to the chemically active site due to the steric hindrance effects. For this reason, typically a hot-injection method is utilized to obtain nanocrystals of sufficient size by increasing the reaction rate if bulky precursors are utilized.[4,20,21] In accordance with this general rule, we found advantageous to utilize n-octylamine as a shorter substitute for oleylamine to form the alkyl ammonium (poly)selenide precursor. In this synthesis route, n-octylamine acts both as the reaction medium for the solvation of the precursors and as a stabilizing agent for $AgBiSe_2$ CNCs, similar to the method utilized for $AgBiS_2$ CNCs synthesized under similar conditions.[38]

Our experimental findings demonstrated that $SeO_2$ could be reduced and dissolved successfully using a mixture of 1-octanethiol (RSH) and 1-octylamine ($RNH_2$) at room temperature following the overall reaction (assuming complete reduction of $Se^{4+}$ to $Se^{2-}$) given in Equation 1 (See Figure S1 for visual demonstration).

$$2\ RNH_2 + SeO_2 + 6\ RSH \rightarrow (RNH_3)_2\ Se + 3\ RS-SR + 2\ H_2O \qquad (1)$$

In this reaction, every two molecules of alkyl thiols oxidize and combine to form a dialkyl dithiol molecule in the presence of alkyl amines and $SeO_2$, yielding an electron pair in the process to reduce $Se^{4+}$, similar to the results presented in a previous report.[4] After the reduction of $Se^{4+}$, the resulting selenide ions are dissolved by complexing with alkyl amine molecules and form the main selenium precursor of the reaction. After the formation of the Se precursor, the presence of AgI and $BiI_3$ in the reaction mixture triggers the formation of $AgBiSe_2$ CNCs in a simplified reaction route given in Equation 2.

$$AgI + BiI_3 + 2\ (RNH_3)_2 Se \rightarrow AgBiSe_2 + 4\ RNH_3I \qquad (2)$$

Although this selenium precursor is sufficiently stable in air in the presence of excess amine-thiol solvent, it should be noted that it cannot be stored in air indefinitely due to constant loss of amine-thiol solvent through redox cycle of Se in air.[4] The in-situ formed alkyl ammonium (poly)selenide precursor slowly decomposes in air to form elemental Se precipitate rather than reverting back to white-colored selenium dioxide if no metal cation that can react with the selenium precursor is present in the solution (Figure S2). No foreign peaks, which may result from $SeO_2$, was detected in XRD scans, showing that this solvent mixture was effective enough for the complete reduction of $SeO_2$ and that it was a strong antioxidant for Se at different oxidation states.

As the first part of this study, we synthesized ternary $AgBiSe_2$ CNCs using our ambient synthesis technique. TEM micrographs showed that $AgBiSe_2$ CNCs had spherical shapes with an average size of 6 nm (Figure 1a, more TEM images can be found in Figure S7).

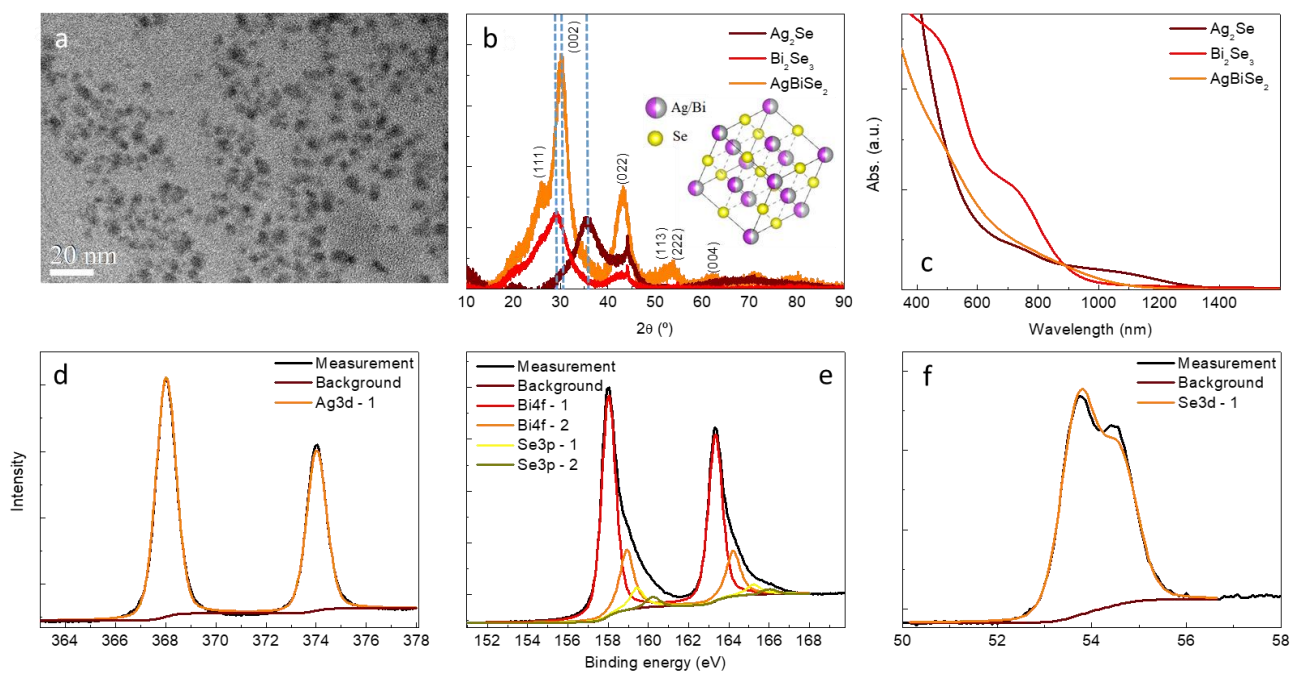

Figure 1 Structural and optical characterization of binary and ternary CNCs of Ag-Bi-Se semiconductor family. (a) TEM images of AgBiSe$_2$ CNCs. (b) XRD scans (Inset: crystal structure of cubic AgBiSe$_2$) and (c) absorption spectra of the ternary and binary phases (The positions of the main peaks of AgBiSe$_2$, Bi$_2$Se$_3$ and Ag$_2$Se are marked with blue dashed lines.). (d) Ag-3d, (e) Bi-4f, (f) Se-3d XPS spectra of AgBiSe$_2$ CNCs.

X-ray diffraction (XRD) was utilized both to confirm the formation and to demonstrate the phase purity of AgBiSe$_2$ nanocrystals. We also synthesized Ag$_2$Se and Bi$_2$Se$_3$ CNCs as binary reference materials to verify the absence of these binary phases in the final AgBiSe$_2$ CNC dispersion. Our results indicated that only the ternary phase was obtained with a quantity detectable by XRD in the presence of both silver and bismuth precursors (Figure 1b). This finding was also supported by absorption spectroscopy as given in Figure 1c. The absorption spectra of both of the binary phases showed distinct features. For Bi$_2$Se$_3$ CNCs, an excitonic-like peak was detected around 720 nm. Ag$_2$Se CNCs showed two broad peaks located close to

730 and 1060 nm. On the other hand, AgBiSe$_2$ CNCs did not exhibit any peak-like feature in the absorption measurements, which is very similar to the absorption spectrum of AgBiS$_2$ CNCs, as reported previously.[38–41] XPS analysis of the AgBiSe$_2$ CNCs revealed that the sample contains all three constituent elements (Ag, Bi and Se, see Figure 1d - f). A single Ag-3d doublet was located at ∼ 368 eV, in line with previously reported binding energy of Ag in AgBiSe$_2$.[37] Two Bi-4f doublets were sufficient to fit Bi-4f spectrum (at ∼ 158.0 and ∼ 159.0 eV, respectively). The position of the Se-3d 5/2 peak (54.6 eV, Figure 1f) confirmed that selenium was in 2- state and not oxidized under the conditions that the synthesis and purification steps were carried out as higher oxidation states would result in higher binding energies for Se.[37,42] The evaluation of the XPS data showed that the elemental composition of AgBiSe$_2$ CNCs was 1.00:1.02:1.05 (Table 1). A SEM investigation of the deposited films demonstrated that the films were smooth and crack-free (Figure S3). This is an important factor to be considered in nanocrystal device fabrication as any cracks and pinholes can severely affect the performance of the resulting devices. The cross sectional SEM images also revealed the nanocrystalline nature of the films.

Table 1 Elemental analysis of AgBiSe$_2$ CNCs synthesized via metal iodide salts and SeO$_2$ (Normalized to Ag-3d).

| Ag | Bi | Se |
|---|---|---|
| 1.00 | 1.02 | 1.05 |

UPS measurements were performed to determine the energy levels of AgBiSe$_2$ CNCs. It was found that the valence band and Fermi level of these CNCs were located around -4.92 and -4.16 eV, respectively (Figure S4a). According to these energy level values, P3HT was chosen as the electron blocking layer for the use of an electron blocking layer with a similar valence band energy/highest occupied molecular orbital (HOMO) level can provide an efficient hole transport channel between AgBiSe$_2$ CNC layer and the anode of the solar cell, yielding higher PCE for AgBiSe$_2$ CNC solar cells (Figure 2a). The effect of an alternative polymer with a deeper HOMO level (-4.8 versus -5.3 eV for P3HT and PTB7, respectively)[41] on AgBiSe$_2$ CNC solar cells can be found in Table S1.

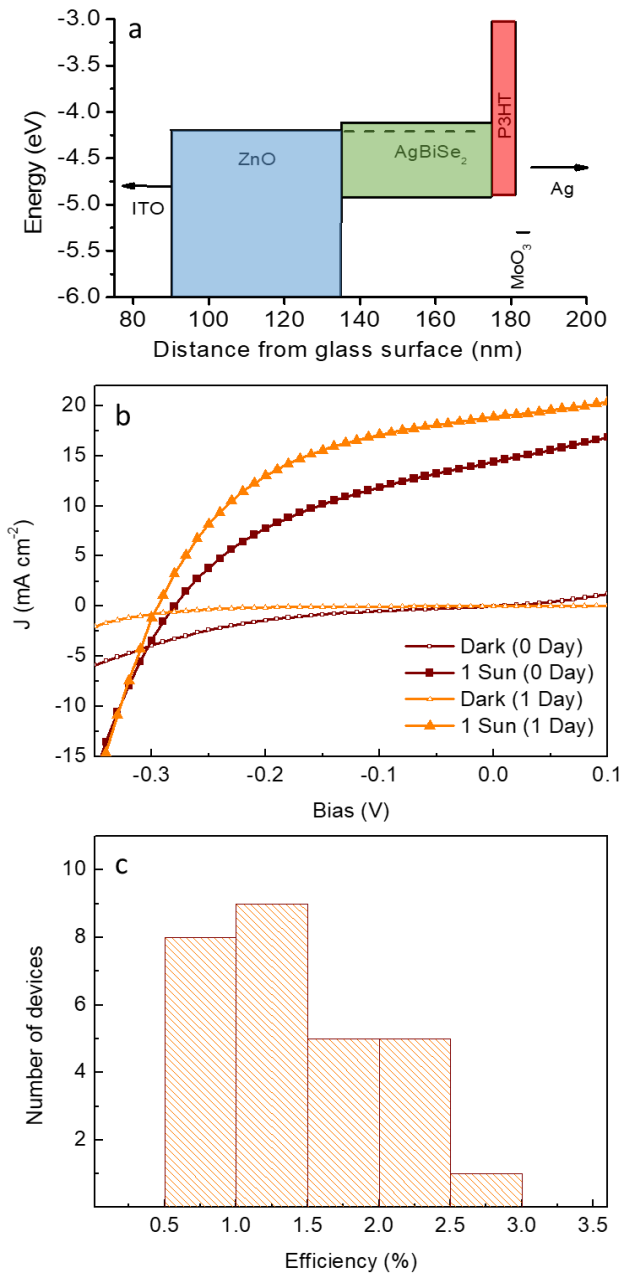

Figure 2 (a) The energy level alignment in the proposed device structure. (b) J-V measurements of the solar cell with a 40 nm-thick AgBiSe$_2$ CNC layer just after metal deposition and after 1-day storage in air. (c) Performance statistics of solar cell with 40 nm-thick AgBiSe$_2$ CNC layer after 1-day storage in air.

We fabricated prototype solar cells of AgBiSe$_2$ CNCs with different thickness using the structure given in Figure 2a. The solar cell with a 40 nm-thick AgBiSe$_2$ NC layer exhibited a $V_{OC}$ of 0.28 V, a $J_{SC}$ of 14.37 mA cm$^{-2}$ and a FF of 0.40 just after fabrication. After 1 day of storage in air, $J_{SC}$ showed a significant increase and reached to 18.88 mA cm$^{-2}$ while $V_{OC}$ increased slightly to 0.29 V. As another observation, the J-V curve demonstrated a dramatic change because of FF increasing from 0.40 to 0.47 after 1-day storage in air (Figure 2b and Table 2), pushing PCE to 2.6 %. This behavior, the increase of efficiency after exposure to air, is not rare and has been reported for different types of CNC solar cells,[43,44] including AgBiS$_2$ CNC solar cells.[38,41] To the best of our knowledge, this is the highest efficiency reported in the literature for any solar cell based on AgBiSe$_2$ CNCs. The average PCE of solar cells with 40-nm thick AgBiSe$_2$ CNC layer was measured to be 1.40 ± 0.55 % after 1-day storage in air (Figure 1c). EQE curves of the solar cells with different AgBiSe$_2$ CNC layer thickness can be found in Figure S4b.

Table 2 Performance of solar cells of AgBiSe$_2$ CNCs just after metal deposition and after 1-day storage in air.

| Time in Air (Days) | $V_{OC}$ (V) | $J_{SC}$ (mA cm$^{-2}$) | FF | Efficiency (%) | Thickness of AgBiSe$_2$ (nm) |
|---|---|---|---|---|---|
| 0 | 0.24 | 10.57 | 0.43 | 1.08 | 20 |
| 0 | 0.28 | 14.37 | 0.40 | 1.59 | 40 |
| 0 | 0.27 | 13.49 | 0.37 | 1.34 | 60 |
| 1 | 0.25 | 12.52 | 0.46 | 1.43 | 20 |
| 1 | 0.29 | 18.88 | 0.47 | 2.60 | 40 |
| 1 | 0.32 | 11.17 | 0.46 | 1.65 | 60 |

In this study, we also investigated the possibility of synthesizing alloyed quaternary AgBiS$_x$Se$_y$ CNCs under ambient conditions. This is an important part of this work, both as alloying in Ag-Bi-S-Se system is not investigated in nanocrystal form in depth yet, except very few reports,[45] and also as it will give a means to tune the bandgap of these CNCs due to different bandgaps of the AgBiS$_2$ and AgBiSe$_2$. We utilized a mixed anion precursor approach for the synthesis of the quaternary CNCs. For both sulfur and selenium can be dissolved in alkyl amines via formation of a amine-chalcogenide complex, we found beneficial to employ alkyl amines as the reaction medium for the synthesis of alloyed AgBiS$_x$Se$_y$ CNCs. The elemental analysis of the AgBiS$_x$Se$_y$ CNCs showed that the selenium ratio in the precursor solution had a direct effect on the Se/S ratio of the resulting CNCs (Figure 3). The presence of both selenium and sulfur was verified with the presence of both Se-3d and S-2s/Se-3s peaks in the respective parts of the XPS spectrum (Figure 3). The comparative analysis of the peak(s) present in these spectra demonstrated that Se/S ratio of the CNCs was in good agreement with the Se/S ratio of the precursor solution (Figure 3f). For XPS quantification, the S-2s/Se-3s spectra of the all samples were fitted using one and two Gaussian - Lorentzian mixed curves for pure (AgBiS$_2$ and AgBiSe$_2$) and alloyed CNCs (AgBiS$_x$Se$_y$), respectively. By examining XPS spectra, it can be deduced that the peak located at ∼ 226 eV originates from S-2s orbital whereas the peak at ∼

229 eV results from Se-3s orbital. Moreover, similar to the case of pure AgBiSe$_2$ CNCs, the Se-3d 5/2 peak of AgBiS$_x$Se$_y$ CNCs located at 54.6 eV confirmed that selenium was in 2- state and not oxidized under the conditions that the synthesis and purification steps were performed.

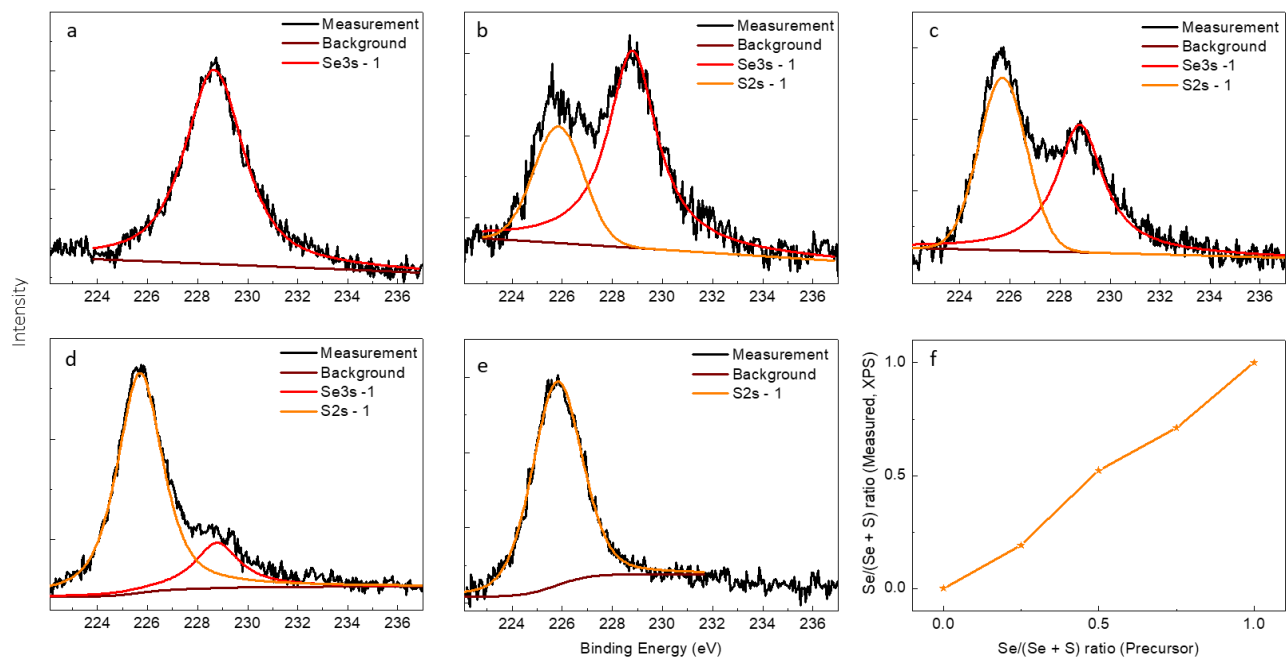

Figure 3 S-2s/Se-3s spectra of AgBiS$_x$Se$_y$ nanocrystals for varying amounts of Se/(Se + S) ratios. (a) 100 %, (b) 75 %, (c) 50 %, (d) 25 % and (e) 0%. (f) The effect of Se/(Se+S) precursor ratio on the Se/(Se+S) ratio of the resulting CNCs.

Ag-3d doublet of the samples was positioned around 368 eV, similar to the AgBiSe$_2$ samples (Figure S5). The change in Se ratio in these samples can also be confirmed via side-by-side analysis of Bi-4f spectra as both S-2p and Se-3p doublets are located within Bi-4f XPS window. The Se-3d doublet of the samples showed the same trend with varying precursor Se ratio. Thus, it can be concluded that the Se content of the resulting product can be tuned by controlling the selenium content of the precursor solution. For further comparison of the content of the samples, the elemental ratios of the CNCs (normalized to Ag-3d) are also provided in Table 3.

Table 3 Elemental composition of the AgBiS$_x$Se$_y$ CNCS synthesized with precursors of different Se/(Se + S) ratios.

| Se/(Se + S) (Precursor) | Ag (XPS) | Bi (XPS) | Se (XPS) | S (XPS) |
|---|---|---|---|---|
| 75 % | 1.00 | 0.89 | 0.69 | 0.28 |
| 50 % | 1.00 | 0.95 | 0.56 | 0.52 |
| 25 % | 1.00 | 0.90 | 0.28 | 1.17 |

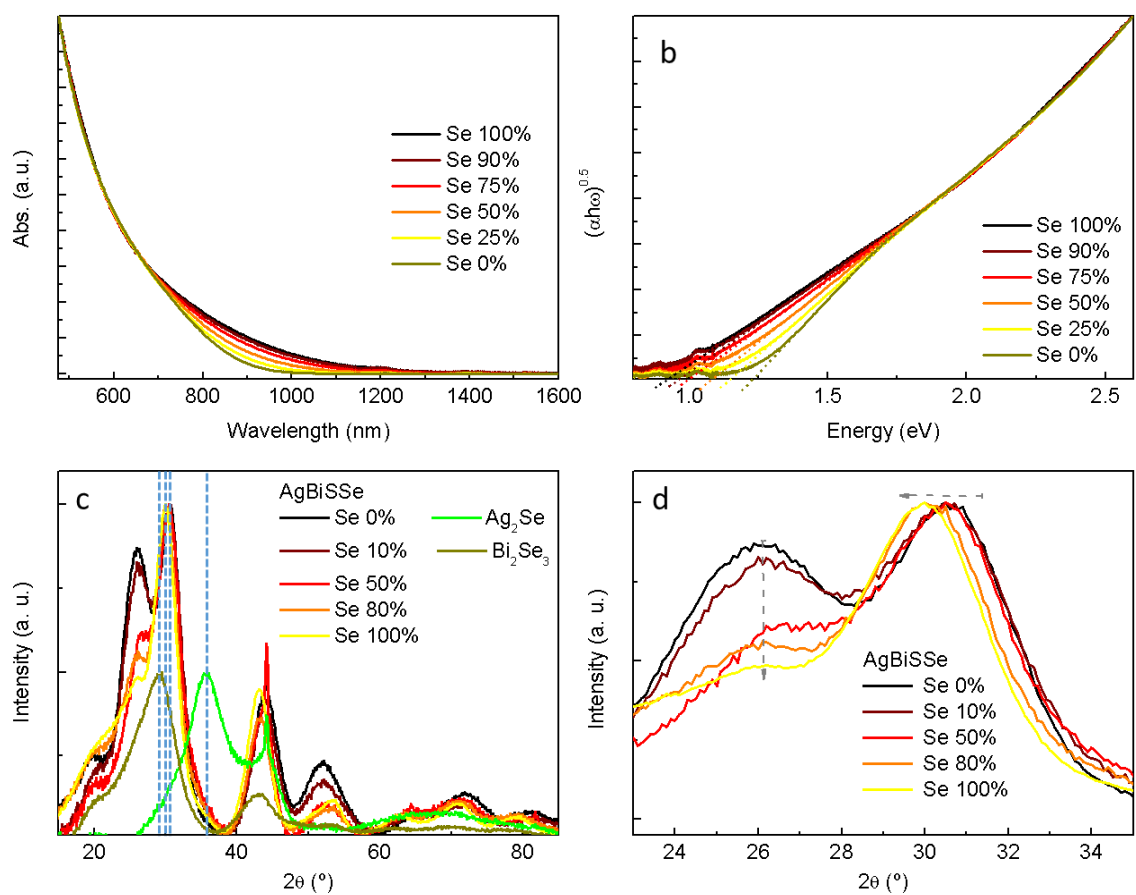

Figure 4 Optical and structural analysis of AgBiS$_x$Se$_y$ CNCs with respect to Se content. (a) Absorbance spectra and (b) Tauc plots of AgBiS$_x$Se$_y$ CNCs. (c) XRD spectra of AgBiS$_x$Se$_y$ CNCs for various values of Se content (The positions of the main peaks of AgBiS$_2$, AgBiSe$_2$, Bi$_2$Se$_3$ and Ag$_2$Se are marked with blue dashed lines.). (d) A magnified version of (c) showing

the shifting/diminishing of the two most prominent peaks of AgBiS$_x$Se$_y$ CNCs for various values of Se content.

After verification of the presence of both S and Se in the CNCs, we used absorption spectroscopy to show the bandgap tunability of the AgBiS$_x$Se$_y$ CNCs. It was observed that the absorbance of CNCs in the 700-1200 nm region increased when the selenium/sulfur ratio of the nanocrystals increased (Figure 4a). The Tauc plots of these alloyed AgBiS$_x$Se$_y$ CNCs yielded a bandgap tuning range of 0.3 eV from 0.92 to 1.23 eV via alloying (Figure 4b). This is an important result for this material family as the bandgap tuning for pure AgBiS$_2$ or AgBiSe$_2$ CNCs was found nontrivial via quantum confinement effect unlike other semiconducting CNCs, such as CdSe, CdTe, HgTe and PbS CNCs.[46–49] Thus, this bandgap tuning strategy can pave the way for better optimization of the optoelectronic properties of these CNCs to enhance the performance of the photovoltaics and thermoelectrics technologies based on AgBiS$_x$Se$_y$ CNCs.

We sought to verify the alloying via alternative methods. As XRD can directly give information about the physical ordering of the atoms in a material, we utilized this method to observe the change in interplanar distances that can result from partial replacement of S with Se. Because of the larger ionic size of selenium compared to that of sulfur, the presence of selenium in the crystal lattice causes an expansion of the lattice and increases the respective lattice parameters. Hence, it becomes possible to determine the sulfur/selenium ratio comparatively in alloyed nanocrystals using the shift of the peaks in the XRD spectra.[50–52] According to Bragg equation (Equation 3),[53] it is expected for higher Se/S ratio to yield smaller Bragg angles. Here, $\lambda$ is the wavelength of the X-ray beam (0.15046 nm for Cu K$\alpha$ radiation), $\theta$ is the Bragg angle, and n is an integer number and d is the lattice spacing.

$$d = \frac{n\lambda}{2\sin(\theta)} \tag{2}$$

XRD analysis demonstrated that the peak located around 30° shifts towards lower angles with increasing Se content, as expected due to the larger ionic size of $Se^{2-}$ with respect to $S^{2-}$ (Figure 4c-d). In addition, the peak located around 26° decreases gradually with the incorporation of more selenium into CNCs as this peak is strong for pure $AgBiS_2$ yet it has a very small amplitude for pure $AgBiSe_2$.[20,45] For pure $AgBiS_2$ CNCs, the interplanar distance was measured to be 0.284 nm using Equation 3, which is very close to the value reported in the literature (0.282 nm, AMCSD 0009219). For pure $AgBiSe_2$ CNCs, this value was found to be 0.291 nm, which matches well with the values documented in previous studies.[54] In addition, no traces of binary products was detected in the XRD spectra of the alloyed $AgBiS_xSe_y$ CNCs for any Se/S ratio. This is in direct accordance with our previous findings on $AgBiS_2$ and $AgBiSe_2$ CNCs. As stated before, although it is possible to form binary products using the same synthesis procedure, these binary products are unlikely to form if both cations are present simultaneously in the synthesis mixture. TEM investigation of the samples demonstrated the formation of nanocrystals for all selenium ratios utilized in this study (Figure 5). High-resolution TEM images of the nanocrystals confirmed that the ambient condition synthesis yielded nanocrystalline material instead of amorphous nanoparticles. The presence of sharp peaks in the SAED image verified the formation of nanocrystalline phase, which is in support of the results obtained by XRD analysis.

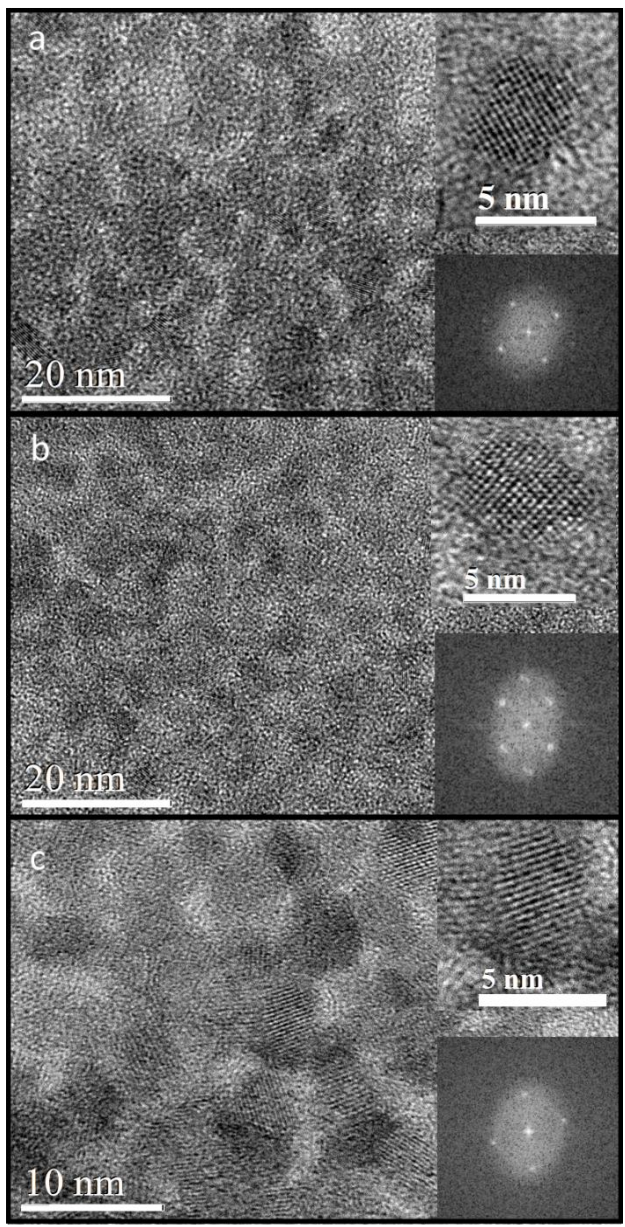

Figure 5 TEM images of AgBiS$_x$Se$_y$ CNCs for different Se/(Se + S) ratios of (a) 75 %, (b) 50 % and (c) 25 %. Se (Insets: upper insets are HRTEM scans whereas lower insets show the SAED spectra of the individual nanocrystals.).

The elemental mapping of AgBiS$_x$Se$_y$ CNCs with different Se/(Se +S) ratios were carried out using a scanning transmission electron microscope (STEM) equipped with EDS. Elemental mapping images are provided in Figure S5 where the EDS signals of S and Se are superimposed on the dark field image to show the distribution of the anions throughout the samples. The results of EDS elemental mapping indicate that the distribution of anions in the samples are uniform and no phase segregation is present, which is in line with XRD results.

**Conclusions**

As a summary, we have demonstrated a simple yet effective room-temperature synthesis method for the environmentally friendly $AgBiSe_2$ nanocrystals that can be easily performed under ambient conditions using alkyl amine - alkyl thiol chemistry. In addition, we have also shown the first solution-processed solar cell based on $AgBiSe_2$ CNCs. The fabricated devices show a preliminary power conversion efficiency of 2.6 %. Furthermore, we have provided a detailed study on the synthesis of alloyed $AgBiS_xSe_y$ CNCs and showed that the bandgap of these CNCs can be tuned by simply changing the Se/S ratio. The formation of quaternary $AgBiS_xSe_y$ CNCs has been verified via absorption and XRD measurements. This work paves the way towards photovoltaics-quality absorbers that are cheaper to produce and environmentally friendly from both a material and production perspectives and address the regulatory concerns and synthetic cost of colloidal nanocrystals based on Schlenk line approaches.

**Acknowledgements**

The authors acknowledge financial support from the European Research Council (ERC) under the European Union's Horizon 2020 research and innovation programme (grant agreement no. 725165), the Spanish Ministry of Economy and Competitiveness (MINECO), and the "Fondo Europeo de Desarrollo Regional" (FEDER) through grant TEC2017-88655-R. The authors also acknowledge financial support from Fundacio Privada Cellex, the program CERCA and from the Spanish Ministry of Economy and Competitiveness, the ICFOstepstone - PhD Programme for Early-Stage Researchers in Photonics, funded by the Marie Skłodowska-Curie Co-funding of regional, national and international programmes (GA665884) of the European Commission, through the "Severo Ochoa" Programme for Centres of Excellence in R&D (SEV-2015-0522).

## Supporting Information

The details of the synthesis, material and device characterization methods, XRD of the decomposed Se precursor, SEM and UPS of $AgBiSe_2$ CNCs, efficiency and EQE spectra of $AgBiSe_2$ CNC solar cells, and XPS/EDS of $AgBiS_xSe_y$ CNCs.

## Notes

The authors declare no competing financial interest.

**TOC Image**

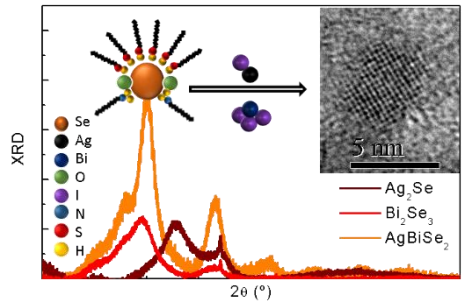